\documentclass[aps,prb,twocolumn,a4paper,superscriptaddress,amsmath,amssymb]{revtex4}

\usepackage[dvips]{graphicx}
\usepackage[dvips]{color} 
\usepackage{bbold}
\usepackage{subfigure}

\newcommand{\trace}[1]{\text{Tr}\left\{ #1\right\}}
\renewcommand{\vec}[1]{\mathbf #1}
\newcommand{\imag}{{\mathrm i}}

\newcommand{\pdiff}[2]{\frac{\partial #1}{\partial #2}}
\newcommand{\pdiffdiff}[2]{\frac{\partial^2 #1}{\partial #2^2}}
\renewcommand{\d}{\mathrm d}

\renewcommand{\Im}[1]{\text{Im}\left\{ #1\right\}}
\renewcommand{\Re}[1]{\text{Re}\left\{ #1\right\}}

\renewcommand{\hbar}{\hslash}
\newcommand{\unitmatrix}{\mathbb{1}}

\begin{document}

\title{Circuit theory of crossed Andreev reflection}

 \author{Jan Petter Morten}
 \email{jan.morten@ntnu.no}
 \affiliation{Department of Physics, Norwegian University of Science
   and Technology, N-7491 Trondheim, Norway}
 
 \author{Arne Brataas}
 \affiliation{Department of Physics, Norwegian University of Science
   and Technology, N-7491 Trondheim, Norway}
 \affiliation{Centre for
   Advanced Study, Drammensveien 78, Oslo, N-0271 Norway}

 \author{Wolfgang Belzig}
 \affiliation{University of Konstanz, Department of Physics, D-78457
   Konstanz, Germany}

 \date{9 September, 2006}

 \begin{abstract}
   We consider transport in a three terminal device attached to one
   superconducting and two normal metal terminals, using the circuit
   theory of mesoscopic superconductivity. We compute the nonlocal
   conductance of the current out of the first normal metal terminal
   in response to a bias voltage between the second normal metal
   terminal and the superconducting terminal. The nonlocal conductance
   is given by competing contributions from crossed Andreev reflection
   and electron cotunneling, and we determine the contribution from
   each process. The nonlocal conductance vanishes when there is no
   resistance between the superconducting terminal and the device, in
   agreement with previous theoretical work. Electron cotunneling
   dominates when there is a finite resistance between the device and
   the superconducting reservoir.  Dephasing is taken into account,
   and the characteristic timescale is the particle dwell time. This
   gives rise to an effective Thouless energy. Both the conductance
   due to crossed Andreev reflection and electron cotunneling depend
   strongly on the Thouless energy. We suggest to experimentally
   determine independently the conductance due to crossed Andreev
   reflection and electron cotunneling in measurement of both energy
   and charge flow into one normal metal terminal in response to a
   bias voltage between the other normal metal terminal and the
   superconductor.
 \end{abstract}
 
 \pacs{} \keywords{}

\maketitle

\section{\label{sec:intro}Introduction}
Crossed Andreev reflection\cite{byers:306,deutsher:apl00} transforms
an incident electron from one conductor, attached to a superconductor,
into a hole in a geometrically separated second attached conductor. In
an alternative, equivalent picture, two quasiparticles from two
separate conductors are transferred into a superconductor as a Cooper
pair.  Electrons can also be transferred between the conductors by
electron cotunneling, where an incident electron tunnels via a virtual
state in the superconductor. The nonlocal conductance, defined in a
three terminal device as the current response in one normal metal
terminal to a voltage bias between the other normal metal and the
superconductor, is determined by contributions from both crossed
Andreev reflection and electron cotunneling. Crossed Andreev
reflection and electron cotunneling give opposite contributions to the
nonlocal conductance.  In this way, crossed Andreev reflection
competes with electron cotunneling.  The realization of a system where
crossed Andreev reflection can be observed, has been the aim of both
experimental\cite{beckmann:prl04,russo:prl05} and
theoretical\cite{Falci:epl01,yamashita:prb03-174504,sanchez:214501,chtchelkatchev:jetp03,giazotto:087001}
work lately. This interest is due to the fact that crossed Andreev
reflection is an inherently mesoscopic phenomenon, with the prospect
of creating entangled
electrons.\cite{recher:165314,nikolai:161320,recher:267003}

The nonlocal conductance of a device where two normal conductors are
tunnel coupled to a bulk superconductor was calculated in second order
perturbation theory for quantum tunneling\cite{hekking:prb94} in Ref.
\onlinecite{Falci:epl01}. The conductance originating from crossed
Andreev reflection was predicted to exactly cancel the conductance due
to electron cotunneling.  Subsequently, disorder\cite{feinberg:epjb03}
and higher order quantum interference
effects\cite{melin:174509,melin:174512} have been incorporated into
this approach, and the noise and cross correlations have been
considered.\cite{bignon:epl04} Ferromagnetic contacts were also
considered in Ref. \onlinecite{Falci:epl01}.  Crossed Andreev
reflection is favored in an antiparallel configuration of the
magnetizations, since Cooper pairs in singlet superconductors consist
of two electrons with opposite spin.  Electron cotunneling is favored
in a parallel configuration.

The predicted dependence of the nonlocal conductance on the
magnetization configuration was observed experimentally in a hybrid
superconducting/ferromagnetic device.\cite{beckmann:prl04}
Subsequently, a bias dependent nonlocal conductance was observed in a
more complicated geometry with only {\it normal} metal contacts to the
superconductor.\cite{russo:prl05} For bias voltages corresponding to
energies below the Thouless energy associated with the distance
between the two normal terminals, a nonlocal signal with sign
corresponding to electron cotunneling was seen. Thus, in contrast to
the results of Refs. \onlinecite{Falci:epl01} and
\onlinecite{feinberg:epjb03}, experiments showed a finite nonlocal
conductance at low bias. Additionally, the sign of the nonlocal signal
in Ref. \onlinecite{russo:prl05} changes when the bias voltage exceeds
the Thouless energy, and this was interpreted as a consequence of
crossed Andreev reflection dominating the nonlocal signal. These
experimental findings are currently not understood.

In previous theoretical works, it is assumed that superconducting
properties, e.g. the magnitude of the gap, are not modified by the
presence of the contacts. This assumption is valid as long as the
coupling between the normal/ferromagnetic conductor and the
superconductor is weak or has a small cross section compared to the
superconducting coherence length.  None of the mentioned theoretical
works describe a dependence in the conductances on the Thouless
energy.

The circuit theory of mesoscopic
transport\cite{nazarov:prl94,nazarov:sm99} is a suitable framework to
understand transport properties of mesoscopic small normal
metal/superconducting systems. Circuit theory is a discretization of
the quasiclassical theory of superconductivity,\cite{Schmid:jltp75}
and can treat nonequilibrium effects and dephasing. A circuit is
modeled as a network of cavities, connectors, and terminals - similar
to the way we understand classical electric circuits.  Terminals are
voltage probes in local thermodynamic equilibrium, whereas cavities
can be driven out of equilibrium.  Cavities and terminals may be
normal metals or superconductors. The connectors can represent
physical interfaces between cavities and terminals or model diffusion.
Connectors representing interfaces are described by their sets of
transmission probabilities.  ``Kirchhoff's rules'' determine the
matrix Green's functions (potentials) of the cavities and the matrix
currents through the connectors. The matrix currents describe not only
the flow of charge and energy, but e.g.  also the flow of
quasiparticle correlation. Circuit theory has been applied
successfully to explain various phenomena in superconductor and normal
metal/ferromagnet hybrid structures, like the proximity
effect,\cite{nazarov:prl94} multiple Andreev
reflections,\cite{bezuglyi:prb00} spin
transport,\cite{Belzig:prb00,Huertas-Hernando:prl02} and
unconventional superconductivity.\cite{tanaka:167003} Circuit theory
has also proved to be a successful approach to calculate the full
counting statistics of hybrid
structures.\cite{belzig:197006,belzig:067006,vanevic:prb05,borlin:197001}
A circuit theory of magnetoelectronics has been developed for hybrid
systems of ferromagnets and normal metals, see Ref.
\onlinecite{brataas:physrep06} and references therein.

We apply circuit theory to calculate the nonlocal conductance of a
three terminal device with contacts to one superconducting and two
normal metal terminals. We substantially generalize previous
theoretical approaches by computing analytically the nonlocal
conductance with general connectors ranging from e.g. ballistic point
contacts via diffusive contacts to tunnel junctions. We take the
proximity effect into account, in which superconducting correlations
affect the spectral properties of a normal metal. We also take
dephasing into account, where the dwell time gives rise to an
effective Thouless energy. We consider low bias transport so that the
relevant energies are much smaller than the superconducting gap.  The
model we consider has a simple and generic geometrical structure.  We
do not consider Josephson effects.  We recover several aspects seen in
experiments.\cite{russo:prl05} Crossed Andreev reflection and electron
cotunneling do not cancel each other.  However, in the limiting case
of strong coupling between the device and the superconducting
terminal, our results agree with previous theoretical work, and the
nonlocal conductance vanishes. The differential conductances depend on
the effective Thouless energy. A dependence on the Thouless energy has
been observed experimentally.\cite{russo:prl05} However, we do not
find an agreement with the sign of the measured nonlocal conductance
above the Thouless energy for the simple model we study.

The paper is organized in the following way: In
Section~\ref{sec:model} we give an overview of our model and the
circuit components. In Section~\ref{sec:ct} we present the
mathematical description and calculation that determines the
conductances associated with the various transport processes in the
system. We show numerical results for some experimentally relevant
systems in Section \ref{sec:results}. Finally, we give our conclusions
in Section~\ref{sec:conclusion}.

\section{\label{sec:model}Model}
\begin{figure}[htbp]
  \centering
  \begin{picture}(0,0)%
\includegraphics{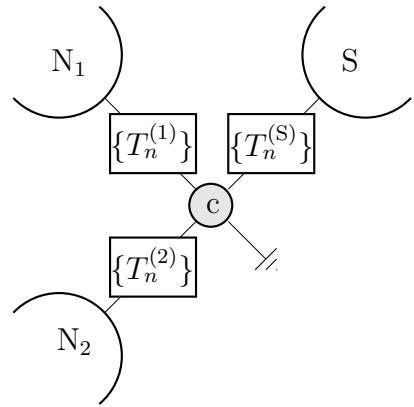}%
\end{picture}%
\setlength{\unitlength}{1776sp}%
\begingroup\makeatletter\ifx\SetFigFont\undefined%
\gdef\SetFigFont#1#2#3#4#5{%
  \reset@font\fontsize{#1}{#2pt}%
  \fontfamily{#3}\fontseries{#4}\fontshape{#5}%
  \selectfont}%
\fi\endgroup%
\begin{picture}(5596,5596)(503,-6759)
\put(1876,-3263){\makebox(0,0)[lb]{\smash{\SetFigFont{5}{6.0}{\familydefault}{\mddefault}{\updefault}{\color[rgb]{0,0,0}{\large $\{T^{(1)}_n\}$}}%
}}}
\put(1876,-4996){\makebox(0,0)[lb]{\smash{\SetFigFont{5}{6.0}{\familydefault}{\mddefault}{\updefault}{\color[rgb]{0,0,0}{\large $\{T^{(2)}_n\}$}}%
}}}
\put(3526,-3271){\makebox(0,0)[lb]{\smash{\SetFigFont{5}{6.0}{\familydefault}{\mddefault}{\updefault}{\color[rgb]{0,0,0}{\large $\{T^{(\text{S})}_n\}$}}%
}}}
\put(1051,-2086){\makebox(0,0)[lb]{\smash{\SetFigFont{5}{6.0}{\familydefault}{\mddefault}{\updefault}{\color[rgb]{0,0,0}{\large N$_1$}}%
}}}
\put(5101,-2086){\makebox(0,0)[lb]{\smash{\SetFigFont{5}{6.0}{\familydefault}{\mddefault}{\updefault}{\color[rgb]{0,0,0}{\large S}}%
}}}
\put(1126,-5986){\makebox(0,0)[lb]{\smash{\SetFigFont{5}{6.0}{\familydefault}{\mddefault}{\updefault}{\color[rgb]{0,0,0}{\large N$_2$}}%
}}}
\put(3211,-4073){\makebox(0,0)[lb]{\smash{\SetFigFont{5}{6.0}{\familydefault}{\mddefault}{\updefault}{\color[rgb]{0,0,0}{\large c}}%
}}}
\end{picture}
  \caption{Our circuit theory model. A normal metal
    chaotic cavity (c) is connected to one superconducting (S) and two
    normal metal terminals (N$_1$ and N$_2$). The three connectors are
    described by their sets of transmission probabilities. A coupling
    to ground represents the ``leakage of coherence'' (see text).}
  \label{fig:ct}
\end{figure}
We consider a three terminal system where one superconducting terminal
and two normal metal terminals are connected to a small normal metal
cavity. We assume that the cavity is large enough that charging
effects can be neglected, and that the Green's functions are isotropic
due to chaotic scattering. A physical realization of the chaotic
cavity could be a small piece of diffusive metal, embedded in a
circuit by e.g. tunneling contacts to the terminals. The assumptions
on the chaotic cavity are quite general and can also be fulfilled for
e.g. a quantum dot with ballistic point contacts if the conductance of
the contacts is much less than the Sharvin estimation of the cavity
conductance.\cite{nazarov:sm99} The circuit theory representation of
our model is shown in Fig.\ \ref{fig:ct}. The normal terminals N$_1$
and N$_2$, and the superconducting terminal S, are connected to the
chaotic cavity c through general connectors represented by their sets
of transmission probabilities $\left\{T_n^{(i)}\right\}$ where
$i=1,2,\text{S}$, and the index $n$ numbers the conductance channels.
These connectors can represent anything from ballistic point contacts
to tunnel junctions.\cite{Belzig:prb00} For a ballistic connector all
transmission eigenvalues are equal to 1 for the propagating channels
and 0 otherwise. For a tunnel junction, all transmission probabilities
are small. Dephasing is represented in the circuit diagram by a
coupling to ground, although no energy or charge current can flow to
this terminal. The dephasing will be discussed in more detail in
Section \ref{sec:ct}.  Our model has a generic geometrical structure
and will capture the physics of crossed Andreev reflection and
electron cotunneling for a wide range of systems.

Let us now identify the various transport processes in the system. We
expect the following contributions to the current: Electron
cotunneling (EC) between terminals N$_1$ and N$_2$, direct Andreev
reflection (DA) between the superconductor and either normal terminal
N$_1$ or N$_2$, and crossed Andreev reflection (CA) between the
superconductor and both normal metal terminals N$_1$ and N$_2$. In
direct Andreev reflection, an injected particle from one terminal
gives rise to a reflected hole in the same terminal, whereas in
crossed Andreev reflection an injected particle from terminal N$_2$
(N$_1$) gives rise to a reflected hole in terminal N$_1$ (N$_2$).
These processes are illustrated in Fig. \ref{fig:processes}.

Semiclassical probability arguments show that the spectral charge
current in the connector between N$_1$ and c at energy $E$ has the
following structure,\cite{Belzig:prb00}
\begin{widetext}
  \begin{center}
    \begin{figure}[htbp]
      \mbox{ \subfigure[Crossed Andreev reflection: A particle from
        N$_2$ with energy $eV_2$ and a particle from N$_1$ with energy
        $-eV_2$ form a Cooper pair in
        S.]{
          \begin{picture}(0,0)%
\includegraphics{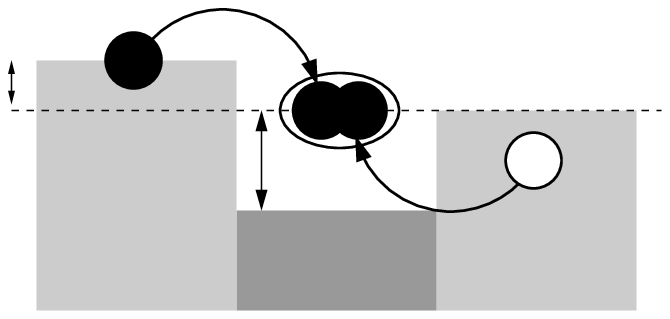}%
\end{picture}%
\setlength{\unitlength}{1579sp}%
\begingroup\makeatletter\ifx\SetFigFont\undefined%
\gdef\SetFigFont#1#2#3#4#5{%
  \reset@font\fontsize{#1}{#2pt}%
  \fontfamily{#3}\fontseries{#4}\fontshape{#5}%
  \selectfont}%
\fi\endgroup%
\begin{picture}(8775,3638)(1,-5762)
\put(2176,-5386){\makebox(0,0)[lb]{\smash{\SetFigFont{5}{6.0}{\familydefault}{\mddefault}{\updefault}{\color[rgb]{0,0,0}{\large N$_2$}}%
}}}
\put(4651,-5386){\makebox(0,0)[lb]{\smash{\SetFigFont{5}{6.0}{\familydefault}{\mddefault}{\updefault}{\color[rgb]{0,0,0}{\large S}}%
}}}
\put(7051,-5386){\makebox(0,0)[lb]{\smash{\SetFigFont{5}{6.0}{\familydefault}{\mddefault}{\updefault}{\color[rgb]{0,0,0}{\large N$_1$}}%
}}}
\put(8776,-3436){\makebox(0,0)[lb]{\smash{\SetFigFont{5}{6.0}{\familydefault}{\mddefault}{\updefault}{\color[rgb]{0,0,0}{\large $E_\text{F}$}}%
}}}
\put(3976,-4111){\makebox(0,0)[lb]{\smash{\SetFigFont{5}{6.0}{\familydefault}{\mddefault}{\updefault}{\color[rgb]{0,0,0}{\large $\Delta$}}%
}}}
\put(  1,-3211){\makebox(0,0)[lb]{\smash{\SetFigFont{5}{6.0}{\familydefault}{\mddefault}{\updefault}{\color[rgb]{0,0,0}{\large $eV_2$}}%
}}}
\end{picture}
        } \qquad \subfigure[Electron
        cotunneling: A particle from N$_2$ at energy $eV_2$ tunnels
        through the cavity c into N$_1$. The density of states in the
        cavity c is suppressed due to the proximity effect from the
        superconducting terminal.]{
          \begin{picture}(0,0)%
\includegraphics{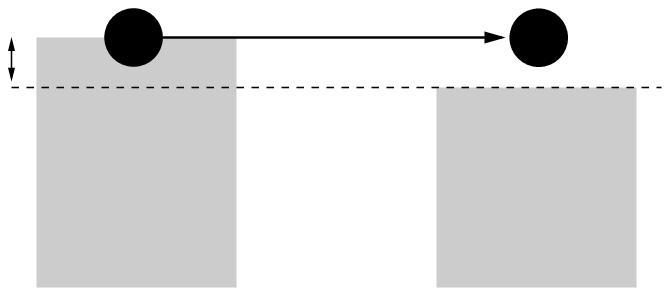}%
\end{picture}%
\setlength{\unitlength}{1579sp}%
\begingroup\makeatletter\ifx\SetFigFont\undefined%
\gdef\SetFigFont#1#2#3#4#5{%
  \reset@font\fontsize{#1}{#2pt}%
  \fontfamily{#3}\fontseries{#4}\fontshape{#5}%
  \selectfont}%
\fi\endgroup%
\begin{picture}(8850,3359)(1,-5762)
\put(2176,-5386){\makebox(0,0)[lb]{\smash{\SetFigFont{5}{6.0}{\familydefault}{\mddefault}{\updefault}{\color[rgb]{0,0,0}{\large N$_2$}}%
}}}
\put(7051,-5386){\makebox(0,0)[lb]{\smash{\SetFigFont{5}{6.0}{\familydefault}{\mddefault}{\updefault}{\color[rgb]{0,0,0}{\large N$_1$}}%
}}}
\put(8851,-3436){\makebox(0,0)[lb]{\smash{\SetFigFont{5}{6.0}{\familydefault}{\mddefault}{\updefault}{\color[rgb]{0,0,0}{\large $E_\text{F}$}}%
}}}
\put(  1,-3211){\makebox(0,0)[lb]{\smash{\SetFigFont{5}{6.0}{\familydefault}{\mddefault}{\updefault}{\color[rgb]{0,0,0}{\large $eV_2$}}%
}}}
\end{picture}
        } }
      \caption{Transport processes in the three terminal device.}
      \label{fig:processes}
    \end{figure}
  \end{center}
\begin{align}
  \label{eq:iE}
  I_1(E)=&\;\frac{G_\text{EC}(E)}{e}\left[f_2(E)-f_1(E)\right]+2\frac{G_\text{DA}(E)}{e}\left[1-f_1(E)-f_1(-E)\right]+\frac{G_\text{CA}(E)}{e}\left[1-f_1(E)-f_2(-E)\right] \, , 
\end{align}
\end{widetext}
where $f_i(\pm E)$ denotes the Fermi-Dirac distribution functions in
normal terminal $i$ at energy $\pm E$. The energy dependent
conductances $G(E)$ are even functions of energy. Andreev reflection
couples an electron with energy $E$ in terminal N$_1$ to an electron
with energy $-E$ in either terminal N$_1$ (DA) or N$_2$ (CA). The
factor 2 for direct Andreev reflection takes into account that two
charges are transmitted in this process. We divide currents and
distribution functions into even and odd parts with respect to energy.
The even part contributes to spectral charge current, and the odd part
contributes to spectral energy current. We therefore construct
$I_\text{T,1}(E)\equiv \left[I_1(E)+I_1(-E)\right]/2$:
\begin{align}
  \label{eq:12}
  I_\text{T,1}(E)=&\frac{1}{e}\left(\frac{1}{2}G_\text{EC}+\frac{1}{2}G_\text{CA}+2G_\text{DA}\right)h_\text{T,1}\nonumber\\
  &-\frac{1}{2e}\left(G_\text{EC}-G_\text{CA}\right)h_\text{T,2},
\end{align}
where $h_{\text{T},i}$ are the reservoir distribution functions
determined by Fermi-Dirac functions.\footnote{The distribution
  functions $h_\text{T}(E),\,h_\text{L}(E)$ can be written in terms of
  the particle distribution function $f(E)$ as
  $h_\text{T}=1-f(E)-f(-E)$ and $h_\text{L}=-f(E)+f(-E)$, see Ref.
  \onlinecite{Belzig:sm99}.} The total charge current is found by
integrating over all energies $I_\text{charge,1}=\int\d E\,
I_\text{T,1}$. We define the nonlocal differential conductance as the
current response in one normal terminal to a voltage between the other
normal terminal and the superconductor. Using Eq. \eqref{eq:12}, this
quantity becomes
\begin{align}
  \pdiff{I_\text{charge,1}}{V_2}=-\int\d E \left[
  G_\text{EC}(E)-G_\text{CA}(E) \right]\pdiff{f(E-eV_2)}{E},
\label{eq:chargecurrent}
\end{align}
where $V_2$ is the voltage in terminal N$_2$. At low temperature, the
derivative of the Fermi function gives a delta-function at energy
$eV_2$. The integral then gives $\partial I_\text{charge,1}/\partial
V_2 = G_\text{EC}(eV_2)-G_\text{CA}(eV_2)$, thus the nonlocal
differential conductance is determined by the difference of the
cotunneling conductance and the crossed Andreev reflection conductance
for quasiparticles at energy $eV_2$. Consequently, a measurement of
the nonlocal conductance does not uniquely determine the conductance
for both processes.

In Section \ref{sec:ct} we show that according to the circuit theory,
$G_\text{EC}-G_\text{CA}$ is positive at all energies for the generic
network considered. This means that the conductance resulting from
electron cotunneling is larger than the conductance resulting from
crossed Andreev reflection, and thus the nonlocal differential
conductance is always positive.

The odd part of the current in Eq.  \eqref{eq:iE} with respect to
energy contributes to energy transport. Direct Andreev reflection does
not contribute since two particles of energies $E$ and $-E$ are
transmitted. The spectral energy current in the connector between
N$_1$ and c has the following structure\footnotemark[\value{footnote}]
\begin{align}
  \label{eq:13}
  I_\text{L,1}(E) \equiv &  \frac{I_1(E)-I_1(-E)}{2}\\\nonumber
  =&\frac{1}{2e}\left(G_\text{EC}+G_\text{CA}\right)\left(h_\text{L,1}-h_\text{L,2}\right).
\end{align}
The total energy current is obtained from the spectral energy current
$I_\text{L,1}$ by $I_\text{energy,1}=\int\d E\,EI_\text{L,1}/e$. The
energy current into the terminal is related by the heat capacity to
the rate of change of the temperature. Thus a nonlocal differential
conductance for energy transport could in principle be measured by
considering the heat flow into terminal N$_1$. We define the nonlocal
differential conductance for energy transport as the energy current
response in one terminal to a voltage between the other normal
terminal and the superconductor. From Eq.  \eqref{eq:13} this quantity
becomes
\begin{align}
  \pdiff{I_\text{energy,1}}{V_2}=-\int\d E\frac{E}{e} \left[
  G_\text{EC}(E)+G_\text{CA}(E) \right]\pdiff{f(E-eV_2)}{E}.
\end{align}
At low temperatures, this gives $\partial I_\text{energy,1}/\partial
V_2 = V_2[G_\text{EC}(eV_2)+G_\text{CA}(eV_2)]$, thus the nonlocal
differential conductance for energy transport is determined by the sum
of the cotunneling conductance and the crossed Andreev reflection
conductance for quasiparticles at energy $eV_2$.

The discussion above shows that the conductance of electron
cotunneling and crossed Andreev reflection can be determined
independently from two experimental quantities. Measurements of the
nonlocal differential conductance for both charge and energy transport
determine the difference and sum of $G_\text{EC}$ and $G_\text{CA}$
respectively. Thus the conductances of electron cotunneling and
crossed Andreev reflection are experimentally accessible and can be
compared to results from theoretical models.

\section{\label{sec:ct}Circuit theory}
The rules of circuit theory allows calculation of the cavity Green's
functions in a network when the terminal Green's functions and
structure of the connectors is determined. The terminals are
characterized by known quasiclassical equilibrium matrix Green's
functions $\check{G}_i$ in Nambu-Keldysh space.\cite{Rammer:rmp86} The
Green's functions depend on quasiparticle energy, and terminal
temperature and chemical potential.  The cavity Green's function in
our model is denoted $\check{G}_\text{c}$. The Green's functions are
4$\times$4 matrices including 2$\times$2 Keldysh space and 2$\times$2
Nambu space. For an explanation of our standard matrix notation, see
Appendix \ref{app:notation}. The general expression for the matrix
current through the connector between terminal $i$ and the cavity
is\cite{nazarov:sm99}
\begin{equation}
  \check{I}_i=-2\frac{e^2}{\pi\hbar}\sum_n
  T^{(i)}_n\left[\check{p}_n^{(i)}\check{G}_i,\check{G}_c\right],
  \label{eq:matrixcurrent}
\end{equation}
where
\begin{equation}
  \check{p}_n^{(i)}=\frac{1}{4+T_n^{(i)}\left(\left\{\check{G}_i,\check{G}_\text{c}\right\}-2\right)}.  
  \label{eq:pni}
\end{equation}
$\check{p}_n^{(i)}$ commutes with the Green's functions in Eq.
\eqref{eq:matrixcurrent} since it can be expanded in anticommutators
of $\check{G}_i$ and $\check{G}_c$.\cite{vanevic:prb05} The matrix
inversion in Eq. \eqref{eq:pni} can be performed analytically in
Keldysh space due to the symmetries of these matrices.  The spectral
charge current can be obtained from the expression for the matrix
current as
$I_{\text{T},i}=\trace{\hat{\sigma}_3\hat{I}_i^\text{(K)}}/8e$, and
the spectral energy current as
$I_{\text{L},i}=\trace{\hat{I}_i^\text{(K)}}/8e$. The K superscript
denotes the Keldysh matrix block of the current.

Correlation between quasiparticles with opposite excitation energy
from the Fermi surface is induced in the cavity due to Andreev
scattering at the superconducting terminal. Cooper pairs transferred
from the superconductor into the cavity give rise to an electron with
excitation energy $E$ and a hole with excitation energy $-E$. The
electron and hole quantum wave functions are initially in phase, but a
relative phase will arise due to a small mismatch of the
wave vectors.\cite{nazarov:sm99} Their wave vectors are
$k=k_\text{F}\sqrt{1\pm E/E_\text{F}}$ where $k_\text{F}$ is the Fermi
momentum and $E_\text{F}$ the Fermi energy. The relevant transport
energy scale is the maximum of the temperature $k_\text{B}T$ and bias
voltage $eV$. The phase difference between the electron and the hole
becomes $\Delta\phi\sim 2E\,\tau/\hbar\approx
2\text{max}(eV,k_\text{B}T)\tau/\hbar$, where the dwell time in the
cavity is $\tau$. The dwell time will be discussed in the next
paragraph. We denote by $E_\text{Th}=\hbar/(2\tau)$ the effective
Thouless energy of the cavity. Let us consider the regime of
negligible dephasing, characterized by $\text{max}(eV,k_\text{B}T)\ll
E_\text{Th}$.  The presence of a superconducting terminal leads to
prevailing electron-hole correlations since $\Delta\phi\sim
~\text{max}(eV,k_\text{B}T)/E_\text{Th}\ll 1$.  In the regime of
complete dephasing, on the other hand, $\text{max}(eV,k_\text{B}T)\gg
E_\text{Th}$, and initial many-particle phase correlation is lost
since $\Delta\phi$ is finite and can only be described statistically.
Thus the induced superconducting correlations due to Andreev
scattering are lost, and the wave functions of the quasiparticles in
the cavity are not in phase. This dephasing effect is described in
circuit theory by an additional terminal for ``leakage of
coherence''.\cite{nazarov:sm99} Note that no charge or energy flows
into this terminal. Circuit theory emerges from a discretization of
the Usadel equation, and the dephasing term stems from the energy term
in this differential equation. The expression for the matrix current
due to dephasing is\cite{nazarov:sm99}
\begin{equation}
  \label{eq:3}
  \check{I}_\text{D}=\imag e^2\nu_0 V_\text{c}E\left[\hat{\sigma}_3 \bar{\unitmatrix},\check{G}_\text{c}\right]/\hbar,
\end{equation}
where $\nu_0$ is the density of states in the normal state and
$V_\text{c}$ is the volume of the cavity.

Let us now discuss the dwell time $\tau$ defined above. The dwell time
can be expressed as $\tau=e^2\nu_0 V_\text{c}R_\text{Total}$.
$R_\text{Total}$ is the total resistance to escape from the system,
and it includes contributions from the contacts and diffusion.
Diffusion is modeled by representing the diffusive region as a network
of cavities connected by tunnel-like conductors with resistance times
area $\tilde{r}=\rho d$. Here $\rho$ is the resistivity and $d$ the
lattice size in the discretized network.\cite{nazarov:sm99} These
connectors contribute to $R_\text{Total}$. When diffusion is the
dominating contribution to $R_\text{Total}$, the definition of the
effective Thouless energy gives $E_\text{Th}=\hbar D/(2L^2)$ in
agreement with the continuum theory from which the circuit theory is
derived. $D$ is the diffusion constant and $L$ the typical length
between contacts.  $E_\text{Th}$ is the relevant energy scale for the
proximity effect in diffusive systems with negligible contact
resistances.\cite{Stoof:prb96} In this paper, however, we will
consider the opposite limit that $R_\text{Total}$ is dominated by the
contact resistances. Spatial variation of the Green's function inside
the system is neglected, and we may discretize with only one cavity.
The effective Thouless energy is in this case
$E_\text{Th}=\hbar/(2e^2\nu_0\tilde{R}L)$, where $\tilde{R}$ is the
sum of the interface resistances in parallel times area.  The contact
resistances induce an energy scale for dephasing similar to systems
where diffusion is the dominant contribution to $R_\text{Total}$. In
Section \ref{sec:results} we show in numerical calculations that the
effective Thouless energy is the relevant energy scale for the
proximity effect in the cavity.

We assume that inelastic processes in the cavity can be neglected
since the characteristic time for inelastic interaction is assumed to
be much larger than the dwell time. The cavity Green's function is
determined by demanding matrix current conservation at each energy.
The sum of all matrix currents flowing into the cavity should vanish,
\begin{equation}
  \label{eq:1}
  \left[-\frac{2e^2}{\pi\hbar}\sum_{\stackrel{\scriptstyle{i=1,2,\text{S}}}{\scriptstyle{n}}}T_n^{(i)}\check{p}_n^{(i)}\check{G}_i+\imag
  \frac{e^2}{\hbar}\nu_0 V_\text{c}E\hat{\sigma}_3 \bar{\unitmatrix},\,\check{G}_\text{c}\right]=0.
\end{equation}
This equation determines the Green's function on the cavity,
$\check{G}_\text{c}$. The retarded and advanced components of
$\check{G}_\text{c}$ can be parameterized in terms of one complex
function $\theta(E)$ as
$\hat{G}^\text{R}_\text{c}=\hat{\sigma}_3\cos(\theta)+\hat{\sigma}_1\sin(\theta)$
and
$\hat{G}^\text{A}_\text{c}=-\hat{\sigma}_3\cos(\theta^*)+\hat{\sigma}_1\sin(\theta^*)$.\cite{Belzig:sm99}
The definition of the Green's function implies that
$\Re{\cos(\theta)}$ is the normalized, energy dependent density of
states in the cavity, $\nu(E)/\nu_0$.

The Keldysh part of the Green's function is parameterized as
$\hat{G}^\text{K}_\text{c}=\hat{G}^\text{R}_\text{c}~\hat{h}_\text{c}-\hat{h}_\text{c}~\hat{G}^\text{A}_\text{c},$
where
$\hat{h}_\text{c}=\hat{\unitmatrix}h_\text{L,c}+\hat{\sigma}_3h_\text{T,c}$.
The normal terminals have Green's functions
$\check{G}_{1(2)}=\hat{\sigma}_3\bar{\tau}_3+(\hat{\sigma}_3h_{\text{L},1(2)}+\hat{\unitmatrix}h_{\text{T},1(2)})(\bar{\tau}_1+\imag\bar{\tau}_2)$,
and the Green's function of the superconducting terminal is
$\check{G}_\text{S}=\hat{\sigma}_1\bar{\unitmatrix}$. We have assumed
that any bias voltage $eV\ll\Delta$, where $\Delta$ is the gap of the
superconducting terminal. Therefore, the only transport process into S
is Andreev reflection since there are no accessible quasiparticle
states in this terminal.

The retarded part of matrix current conservation, Eq. \eqref{eq:1},
gives an equation that determines the pairing angle $\theta$, the
retarded ``Usadel equation'' of the cavity:
\begin{align}
  \label{eq:retardedusadel}
  \left(\imag\frac{e^2}{\hbar}\nu_0 V_\text{c}E
  \right.&\left.-\frac{e^2}{\pi\hbar}\sum_{i=1,2;\,n}\frac{T^{(i)}_n}{2+T^{(i)}_n\left(\cos(\theta)-1\right)}\right)\sin(\theta)\nonumber\\
  &+\frac{e^2}{\pi\hbar}\sum_n\frac{T_n^\text{(S)}}{2+T_n^\text{(S)}\left(\sin(\theta)-1\right)}\cos(\theta)=0.
\end{align}
The physical effect on the spectral properties from the various terms
can be understood by comparing this equation to the corresponding
diffusion equation for a bulk
superconductor.\cite{Schmid:nato81,esteve:nato97,Morten:prb04} This is
given in Eq. (II.29b) of Ref. \onlinecite{Schmid:nato81} and becomes
in our notation\footnote{To make contact with our notation we must
  identify $\alpha=\cos\theta$ and $\beta=\sin\theta$ in Eq.  (II.29b)
  of Ref.  \onlinecite{Schmid:nato81}. In addition to this, we put
  $\gamma=0$ since there is no such component in the retarded Green's
  function of the cavity, $\dot{\Phi}=0$ since we are considering
  stationary phenomena and $\vec{v}_\text{S}=0$ since there is no
  supercurrent in the cavity.}
\begin{equation}
  \label{eq:schmid-retarded}
  \frac{\hbar D}{2}\pdiffdiff{\theta}{x}+\left(\imag E-\frac{\hbar}{2\tau_E}\right)\sin\theta+\Delta\cos\theta-\frac{\hbar}{\tau_\text{sf}}\sin\theta\cos\theta=0,\nonumber
\end{equation}
where $D$ is the diffusion constant, $\Delta$ the gap to be determined
self-consistently, $1/\tau_E$ the inelastic scattering rate, and
$1/\tau_\text{sf}$ the spin-flip scattering rate. Comparing this to
Eq. \eqref{eq:retardedusadel}, we see that the coupling to the
superconductor induces superconducting correlations, and that the
coupling to the normal terminals give quasiparticles a finite
lifetime.  Spin-flip scattering could be included by taking into
account magnetic impurities in the cavity. We consider a normal metal
cavity.  To describe a superconducting cavity, we would have to
include a pairing term in the Hamiltonian. This would result in a term
in Eq.  \eqref{eq:1} with the same structure as the term corresponding
to coupling to the superconducting terminal. As long as we do not
consider Josephson effects, the effect of superconductivity in the
cavity could therefore be included by a quantitative renormalization
of the coupling strength to S, which is straightforward.

In the regime of complete dephasing, the term proportional to
$\sin(\theta)$ dominates in Eq. \eqref{eq:retardedusadel} because of
the large factor $E/E_\text{Th}$. The solution in this limit is
$\theta=0$, which means that there are no electron-hole correlations
in the cavity.

Let us now consider the Keldysh part of Eq. \eqref{eq:1}. We take the
trace of this matrix block after first multiplying it with
$\hat{\sigma}_3$. The resulting equation determines the distribution
function $h_\text{T,c}$ in the cavity,
\begin{align}
  \label{eq:hTc}
  G_\text{T,1}\left(h_\text{T,1}-h_\text{T,c}\right)+G_\text{T,2}&\left(h_\text{T,2}-h_\text{T,c}\right)\nonumber\\
  &+G_\text{T,S}\left(0-h_\text{T,c}\right)=0.
\end{align}
This implies charge conservation at each energy, with effective energy
dependent conductances $G_{\text{T},i}$ between the cavity and
terminals N$_1$, N$_2$ and S. The zero in the term
$G_\text{T,S}\left(0-h_\text{T,c}\right)$ represents the charge
distribution function in the superconductor, $h_\text{T,S}$, which
vanishes since the superconductor is grounded. The conductance
$G_\text{T,S}$ controls the Andreev reflection rate in the cavity
since it is the conductance between the cavity and the superconducting
terminal. The effective conductances are given in terms of the pairing
angle and the transmission probabilities as
\begin{subequations}
\begin{align}
  G_{\text{T},i}&=2\frac{e^2}{\pi\hbar}\sum_nT^{(i)}_n\frac{(2-T^{(i)}_n)\Re{\cos(\theta)}+T^{(i)}_nD_\text{T}}{\left|2+T^{(i)}_n\left(\cos(\theta)-1\right)\right|^2},\\
  G_\text{T,S}&=2\frac{e^2}{\pi\hbar}\sum_nT_n^\text{(S)}\frac{(2-T_n^\text{(S)})\Re{\sin(\theta)}+T_n^\text{(S)}D_\text{T}}{\left|2+T_n^\text{(S)}\left(\sin(\theta)-1\right)\right|^2}.
\end{align}
\label{eq:9}
\end{subequations}
Here $D_\text{T}=\left[\Re{\cos(\theta)}\right]^2+
\left[\Re{\sin(\theta)}\right]^2$ and $i=1,2$. The term proportional
to $G_\text{T,S}$ in Eq. \eqref{eq:hTc} describes conversion of
quasiparticles in the cavity into condensate in the superconducting
terminal. There is an analogous term in the Boltzmann equation for a
continuum superconductor,\cite{Morten:prb04,morten:prb05} which
describes conversion between quasiparticles and superconducting
condensate over the coherence length. This analogous term has a
similar dependence on $\theta$. The rate of this conversion is
controlled by $\Delta$ in the continuum case, and by $G_\text{T,S}$ in
the cavity of our discretized theory.

In the regime of complete dephasing $\theta=0$, the conductances to
the normal terminals $G_{\text{T},i}$ coincide with the
Landauer-B\"{u}ttiker formula. $G_\text{T,S}=2e^2\sum_n
(T_n^{(\text{S})})^2/(\pi\hbar(2-T_n^{(\text{S})})^2)$ corresponds to
the Andreev conductance of an N/S interface in a diffusive system,
calculated by Beenakker.\cite{beenakker:12841} Thus the distribution
function $h_\text{T,c}$ can be determined in the regime of complete
dephasing from well-known results by demanding charge current
conservation.

In the tunnel barrier limit, all transmission probabilities are small,
and we can expand to first order in $T_n^{(i)}$ in Eqns. \eqref{eq:9}.
We define $g_i=e^2\sum_nT^{(i)}_n/(\pi\hbar)$ for $i=1,2,\text{S}$.
For the normal terminals ($i=1,2$) we find
$G_{\text{T},i}=g_i\Re{\cos(\theta)}$. $\Re{\cos(\theta)}$ gives the
normalized density of states in the cavity which is under the
influence of the proximity effect. The tunnel conductance to the
superconductor becomes $G_\text{T,S}=g_\text{S}\Re{\sin(\theta)}$
which vanishes when there is complete dephasing $\theta=0$. This is
expected since the Andreev conductance of a tunnel barrier between
incoherent normal and superconducting terminals
vanishes.\cite{Blonder:prb82}

The trace of the Keldysh block of Eq.  \eqref{eq:1} gives an equation
that determines the distribution function $h_\text{L,c}$ in the
cavity,
\begin{equation}
  \label{eq:hLc}
  G_\text{L,1}\left(h_\text{L,1}-h_\text{L,c}\right)+G_\text{L,2}\left(h_\text{L,2}-h_\text{L,c}\right)=0.
\end{equation}
This is energy conservation at each energy, with effective energy
dependent conductances for energy transport $G_{\text{L},i}$. No
energy current can flow through the contact between the cavity and the
superconducting terminal since no net energy is transferred into S by
Andreev reflection. Our calculation is restricted to $E\ll\Delta$, but
in the general case a quasiparticle current which carries energy can
flow into the superconducting terminal for $E>\Delta$. The effective
conductances for energy transport are given in terms of the pairing
angle and the transmission probabilities as
\begin{align}
  \label{eq:11}
  G_{\text{L},i}&=2\frac{e^2}{\pi\hbar}\sum_nT^{(i)}_n\frac{(2-T^{(i)}_n)\Re{\cos(\theta)}+T^{(i)}_nD_\text{L}}{\left|2+T^{(i)}_n\left(\cos(\theta)-1\right)\right|^2}.
\end{align}
Here
$D_\text{L}=\left[\Re{\cos(\theta)}\right]^2-\left[\Im{\sin(\theta)}\right]^2$
and $i=1,2$.

In the tunnel barrier limit we find that
$G_{\text{L},i}=G_{\text{T},i}=g_i\Re{\cos(\theta)}$ for $i=1,2$,
which means that the effective conductance for energy transport and
charge transport into the normal terminals are equal. The conductances
correspond to the usual quasiparticle tunnel conductance in this case.

Eqns. \eqref{eq:retardedusadel}, \eqref{eq:hTc} and \eqref{eq:hLc}
determine the Green's function in the cavity. The charge and energy
currents, $I_{\text{T},i}$ and $I_{\text{L},i}$, between terminal $i$
and the cavity can be calculated once $\check{G}_\text{c}$ is known.
Comparison of Eqns.  \eqref{eq:12} and \eqref{eq:13} with the
expressions for $I_\text{T,1}$ and $I_\text{L,1}$ obtained from
circuit theory, allows us to determine the conductances associated
with the various transport processes:
\begin{subequations}
\begin{align}
  G_\text{DA}(E)=&\;\frac{1}{4}\left(\frac{G_\text{T,1}\left(G_\text{T,2}+G_\text{T,S}\right)}{G_\text{T,1}+G_\text{T,2}+G_\text{T,S}}-\frac{G_\text{L,1}G_\text{L,2}}{G_\text{L,1}+G_\text{L,2}}\right),\\
  G_{\stackrel{\scriptstyle{\text{EC}}}{\scriptstyle{\text{CA}}}}(E)=&\;\frac{1}{2}\left(\frac{G_\text{L,1}G_\text{L,2}}{G_\text{L,1}+G_\text{L,2}}\pm\frac{G_\text{T,1}G_\text{T,2}}{G_\text{T,1}+G_\text{T,2}+G_\text{T,S}}\right).\label{eq:conductances-ECCA}
\end{align}
\label{eq:conductances}
\end{subequations}
These formulas are the main result of our calculation. Eq.
\eqref{eq:conductances-ECCA} shows that $G_\text{EC}-G_\text{CA}$ is
positive. Thus the nonlocal conductance $\partial I_1/\partial V_2$ is
positive. In the limit where the coupling to the superconducting
terminal vanishes, i.e. all $T_n^\text{(S)}\to 0$, only $G_\text{EC}$
remains nonzero and the conductance agrees with the result for a
normal double barrier system.  If the conductance to one of the normal
terminals vanishes, i.e. all $T_n^{(i)}\to 0$ for e.g. $i=2$, only
$G_\text{DA}$ is nonzero. When the coupling between the
superconducting terminal and the cavity is very strong
$G_\text{T,S}\gg G_{\text{T},1},G_{\text{T},2}$, we recover the result
of Ref.  \onlinecite{Falci:epl01} that $G_\text{EC}=G_\text{CA}$,
which means that the nonlocal conductance vanishes since transport by
electron cotunneling is exactly canceled by crossed Andreev
reflection.

To describe a device where spatial variation in a bulk region is
important, a model with several cavities connected in a network is
required. The connectors between cavities represent the intrinsic
resistance due to diffusion, and contribute to $R_\text{Total}$ and
thus to the effective Thouless energy $E_\text{Th}=\hbar/(2e^2\nu_0
V_\text{c}R_\text{Total})$. However, as long as there are no Josephson
currents in the network, the symmetry
$G_\text{EC(CA)}=\alpha+(-)\beta$ of Eqn.
\eqref{eq:conductances-ECCA} persists. Here $\alpha$ and $\beta$ are
positive numbers. This follows since the currents flowing out of the
normal terminals given by circuit theory can be written
$I_{\text{L},i}(E)=C_{\text{L}}^{(i)}\left(h_{\text{L},1}-h_{\text{L},2}\right)$
and
$I_{\text{T},i}(E)=C_{\text{T}1}^{(i)}h_{\text{T},1}-C_{\text{T}2}^{(i)}h_{\text{T},2}$
where $C_\text{L}^{(i)},\,C_{\text{T}1}^{(i)}$ and
$C_{\text{T}2}^{(i)}$ are coefficients. Comparing these expressions to
Eqns. \eqref{eq:12} and \eqref{eq:13}, we see e.g. that
$G_\text{EC}-G_\text{CA}=2eC_{\text{T2}}^{(i)}>0$ for contact $i$,
regardless of the internal structure of the network of cavities. Thus
the sign of the nonlocal conductance is not affected by diffusion or
network geometry. We interpret this result as the consequence of a
symmetry between crossed Andreev reflection and electron cotunneling.
Both processes involve the transfer of quasiparticles through the
contacts to the normal metals and the network between them, but
crossed Andreev reflection also involves Andreev reflection at the
interface to the superconducting terminal. Thus the resistance
limiting crossed Andreev reflection can at minimum be as small as the
resistance for electron cotunneling unless other physical processes
affect these quantities. We believe that an explanation of the
measurements in Ref.  \onlinecite{russo:prl05} requires additional
physical effects not considered here.

\subsection{\label{sec:analytical-limit}Analytically solvable limits}
In this section, we give results for two limits where it is possible
to solve Eq. \eqref{eq:retardedusadel} analytically and obtain simple
expressions for the conductances in Eqns.  \eqref{eq:conductances}.

\subsubsection{\label{sec:incoherent-limit}The regime of complete dephasing}
When $E\gg E_\text{Th}$ there is complete dephasing. In this case the
conductances $G_{\text{T},i}=G_{\text{L},i}=g_i$ for $i=1,2$ agree
with the Landauer-B\"{u}ttiker formula and $G_\text{T,S}$ is
Beenakker's result for the conductance of an N|S interface in a
diffusive system, as noted above. In this limit, the conductances in
Eq.  \eqref{eq:conductances} become
\begin{subequations}
\begin{align}
  G_\text{DA}=&\frac{g_1^2}{4}\frac{G_\text{T,S}}{(g_1+g_2)(g_1+g_2+G_\text{T,S})}\\
  G_{\stackrel{\scriptstyle{\text{EC}}}{\scriptstyle{\text{CA}}}}(E)=&\frac{g_1g_2}{2}\frac{1}{(g_1+g_2)(g_1+g_2+G_\text{T,S})}\nonumber\\&\times\begin{cases}2(g_1+g_2)+G_\text{T,S}&\text{for
      EC}\\ G_\text{T,S}&\text{for CA}\end{cases}.
\end{align}
\end{subequations}
The currents resulting from these expressions using Eq.  (\ref{eq:iE})
can also be calculated from a semiclassical approach by demanding
charge conservation in the cavity using the well-known theory for
incoherent N/S transport. The result for $G_\text{EC}-G_\text{CA}$ was
shown in Ref.  \onlinecite{sanchez:214501}, where also spin polarizing
contacts are employed to give a negative nonlocal conductance. Again,
in the limit of strong coupling to the superconductor,
$g_i/G_\text{T,S}\ll 1$, we see that $G_\text{EC}=G_\text{CA}$ and
thus the nonlocal conductance vanishes.

\subsubsection{\label{sec:tunnel-limit}Tunnel barrier limit}
In the tunnel barrier limit, all transmission probabilities are small.
We expand to first order in $T_n^{(i)}$ in the expressions for the
matrix currents. This corresponds to putting $\check{p}_n^{(i)}\to
1/4$ in Eq. \eqref{eq:matrixcurrent}. The resulting Eq. \eqref{eq:1}
can be solved analytically without resorting to a parametrization of
$\check{G}_\text{c}$.\cite{nazarov:sm99} The solution is of course
equivalent to what we obtain from Eqns.  \eqref{eq:retardedusadel},
\eqref{eq:hTc} and \eqref{eq:hLc} in the same limit. Let us consider
the resulting expressions for the conductances of nonlocal transport
in some limits. At zero energy, we obtain
\begin{subequations}
\begin{align}
  G_\text{DA}&=\frac{g_1^2}{4}\frac{g_\text{S}^2}{[(g_1+g_2)^2+g_\text{S}^2]^{3/2}},\\
  G_\text{EC}&=\frac{g_1g_2}{2}\frac{2\left(g_1+g_2\right)^2+g_\text{S}^2}{[(g_1+g_2)^2+g_\text{S}^2]^{3/2}},\label{eq:ectunnel}\\
  G_\text{CA}&=\frac{g_1g_2}{2}\frac{g_\text{S}^2}{[(g_1+g_2)^2+g_\text{S}^2]^{3/2}}.
\end{align}
\label{eq:E=0-tunnelconductances}
\end{subequations}
These results correspond to completely phase coherent transport. The
full counting statistics for the same system in this regime has been
calculated in Ref. \onlinecite{borlin:197001}.  In that paper, it is
found that the cross correlation noise can have both signs in the
three terminal device.

If there is complete dephasing, the cavity spectral properties are
like those of a normal metal. This gives $G_{\text{T,S}}=0$ since the
Andreev conductance of a tunnel barrier vanishes for incoherent N/S
transport.\cite{Blonder:prb82} Therefore, there are no transport
channels into the superconducting terminal and it is effectively
isolated from the circuit. Only the conductance for electron
cotunneling is nonzero, and we obtain from Eq. \eqref{eq:ectunnel}
\begin{equation}
  \label{eq:5}
  G_\text{EC}=\frac{1}{1/g_1+1/g_2},
\end{equation}
i.e. addition of the conductances between the cavity and both normal
metal terminals in series. This result corresponds to normal state
tunneling between N$_1$ and N$_2$.

\section{\label{sec:results}Results}
In an experimental situation, nonlocal transport can be probed in
measurements of the voltage or current in terminal N$_1$ resulting
from the injection of current through terminal
N$_2$.\cite{russo:prl05,beckmann:prl04} At zero temperature, the
nonlocal differential conductance $\partial I_\text{charge,1}/\partial
V_2$ as a function of $eV_2/E_\text{Th}$ corresponds to
$G_\text{EC}-G_\text{CA}$, see Eq. \eqref{eq:chargecurrent}. This
quantity is given by Eqns. \eqref{eq:conductances-ECCA} and is always
positive within our model. The detailed behavior as a function of
$eV_2$ depends on the pairing angle of the cavity, $\theta$, which is
determined by Eq. \eqref{eq:retardedusadel}. This equation can be
solved numerically, and determines the spectral properties and thus
the nonlocal conductances when the parameters of the model are fixed.
We will show results for three systems with different combinations of
contacts which can be fabricated with state of the art nanotechnology.
The contacts we study are tunnel barriers or metallic contacts. The
systems represent various realizations where our circuit theory model
applies. We show that the nonlocal conductances are very sensitive to
the type of contacts in the system, and have a strong dependence on
the Thouless energy. We have also investigated the effect of inelastic
scattering inside the cavity, and have found no notable qualitative
differences on the conductances as compared to the elastic case, thus
these results are not shown here.

\subsection{\label{sec:tunnel}Tunnel barriers}
\begin{figure}[htbp]
  \includegraphics[scale=0.5]{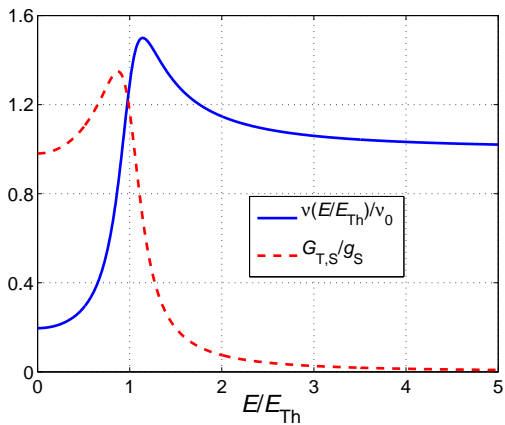}
  \caption{(Color online) Spectral properties of the cavity when all
    contacts are of tunnel type with
    $g_1/g_\text{S}=g_2/g_\text{S}=0.1$. Solid line (blue) shows
    normalized density of states, $\nu(E)/\nu_0$. Dashed line (red)
    shows $G_\text{T,S}/g_\text{S}$ which is the parameter that
    determines the Andreev reflection rate of quasiparticles.}
  \label{fig:spectral-tunnel-symmetric}
\end{figure}

When all the three contacts are tunnel barriers, the equation for
matrix current conservation can be solved analytically as mentioned in
Section \ref{sec:tunnel-limit}. The resulting expressions are quite
complicated. We expand to first order in the transmission
probabilities since $T_n^{(i)}\ll 1$ for $i=1,2,\text{S}$, and let
$g_i$ denote tunnel barrier conductance in the normal metal state. Let
us first consider a symmetric system,
$g_1/g_\text{S}=g_2/g_\text{S}=0.1$. We define the Thouless energy in
this case as $E_\text{Th}=\hbar g_\text{S}/(2e^2\nu_0 V_\text{c})$.
The spectral properties of the cavity are shown in Fig.
\ref{fig:spectral-tunnel-symmetric}.
In this plot we show the normalized density of states in the cavity
and the conductance $G_\text{T,S}$ which controls the Andreev
reflection rate. From Fig. \ref{fig:spectral-tunnel-symmetric} we see
that below the Thouless energy of the cavity, the quasiparticle
density of states is suppressed due to electron-hole correlations.
This affects all transport processes since they rely on quasiparticles
propagating through the cavity. Above the Thouless energy, the density
of states approaches the value in the normal state. This is the
typical behavior for proximity coupled systems.\cite{esteve:nato97}
The conductance between the cavity and the superconducting terminal,
$G_\text{T,S}$, approaches the normal state value at low energies. At
high energies, $G_\text{T,S}$ vanishes, as expected for tunnel
barriers in the regime $E<\Delta$ when the proximity effect can be
neglected.\cite{Blonder:prb82} This affects crossed and direct Andreev
reflection, which vanish when $G_\text{T,S}$ goes to zero.

\begin{figure}[htbp]
  \includegraphics[scale=0.5]{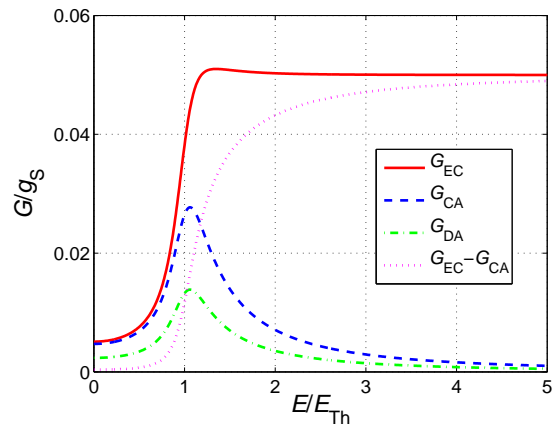}
  \caption{(Color online) Conductance for the various transport
    processes through the cavity when all contacts are of tunnel type.
    The two normal contacts are have the same conductance in this plot,
    $g_1/g_\text{S}=g_2/g_\text{S}=0.1$. Solid line (red)
    $G_\text{EC}/g_\text{S}$, dashed line (blue)
    $G_\text{CA}/g_\text{S}$, dash-dot line (green)
    $G_\text{DA}/g_\text{S}$, and dotted line (purple)
    $(G_\text{EC}-G_\text{CA})/g_\text{S}$.}
  \label{fig:conductances-tunnel-symmetric}
\end{figure}

A plot of the conductances for electron cotunneling, and crossed and
direct Andreev reflection is shown in Fig.
\ref{fig:conductances-tunnel-symmetric}. The conductances have a rapid
increase near $E=E_\text{Th}$ related to the energy dependence of the
density of states. At high energies $G_\text{EC}$ approaches the value
for normal double barrier tunneling, Eq. \eqref{eq:5}. $G_\text{CA}$
and $G_\text{DA}$ vanish at high energy. These conductances are
determined by an interplay of the density of states and the Andreev
reflection rate, and are small when either of these quantities are
small. The measurable nonlocal differential conductance,
$G_\text{EC}-G_\text{CA}$, is a monotonously increasing function with
increasing energy, which starts at small value and approaches
$G_\text{EC}$ above the Thouless energy.

\begin{figure}[htbp]
  \includegraphics[scale=0.5]{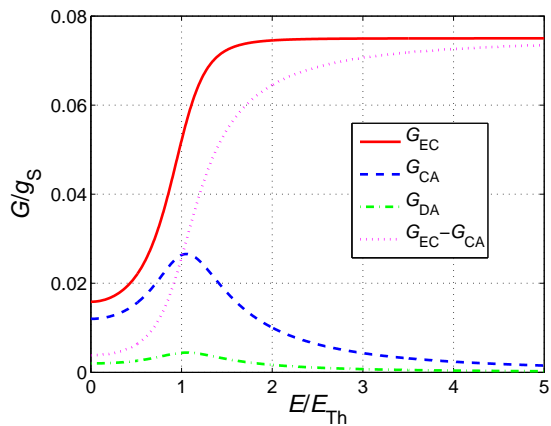}
  \caption{(Color online) Conductance for the various transport
    processes through the cavity when all contacts are of tunnel type.
    Conductances are normalized by $g_\text{S}$, and parameter values
    are $g_1/g_\text{S}=0.1$, $g_2/g_\text{S}=0.3$. Solid line (red)
    $G_\text{EC}/g_\text{S}$, dashed line (blue)
    $G_\text{CA}/g_\text{S}$, dash-dot line (green)
    $G_\text{DA}/g_\text{S}$, and dotted line (purple)
    $(G_\text{EC}-G_\text{CA})/g_\text{S}$.}
  \label{fig:conductances-tunnel}
\end{figure}

Let us now consider the effect of asymmetry between the tunnel
barriers to the normal metals. The expressions for the conductances at
zero energy, Eqns.  \eqref{eq:E=0-tunnelconductances}, show that the
direct Andreev conductance of one contact, $G_\text{DA}^{(1)}$ for
example, is proportional to $g_1^2$ since two electrons have to tunnel
through the connector with conductance $g_1$. $G_\text{DA}^{(1)}$ is,
however, only weakly dependent on $g_2$. The same is true for the
direct Andreev conductance of connector 2 with $g_1\leftrightarrow
g_2$. The direct Andreev conductances are therefore relatively
independent on the asymmetry. On the other hand, nonlocal conductances
are proportional to $g_1g_2$ because a quasiparticle has to tunnel
through both connectors. These conductances are sensitive to the
asymmetry which we define as $a=g_1/g_2$. For the conductance measured
at terminal N$_1$ we see that
$G_\text{EC(CA)}^{(1)}/G_\text{DA}^{(1)}\propto 1/a$.  Conversely, the
conductance measured at terminal N$_2$ gives
$G_\text{EC(CA)}^{(2)}/G_\text{DA}^{(2)}\propto a$. Thus asymmetry
suppresses the nonlocal conductance of one contact, and enhances the
nonlocal conductance of the other contact. In Fig.
\ref{fig:conductances-tunnel} we show the conductances for a system
where $g_1/g_\text{S}=0.1$ and $g_2/g_\text{S}=0.3$. The asymmetry is
now $a=1/3$, thus the effect of nonlocal processes is enhanced when we
consider the conductances measured at terminal N$_1$. The spectral
properties in this case are similar to those of the symmetric system
shown in Fig. \ref{fig:spectral-tunnel-symmetric}, and are not shown
here. Comparing Figs. \ref{fig:conductances-tunnel-symmetric} and
\ref{fig:conductances-tunnel},we see that the conductances for crossed
and direct Andreev reflection are not as peaked in the asymmetric
system as in the symmetric system. We also see that the relative
magnitude of the nonlocal conductances to the direct Andreev
conductance has increased.  In the remainder of the manuscript, we
consider the conductances of contact 1 in asymmetric structures
$a=1/3$ since we are mostly interested in the conductance resulting
from nonlocal processes.

\subsection{\label{sec:Sballistic}Tunnel contacts to N$_1$ and N$_2$,\\
  metallic contact to S}
\begin{figure}[htbp]
  \includegraphics[scale=0.5]{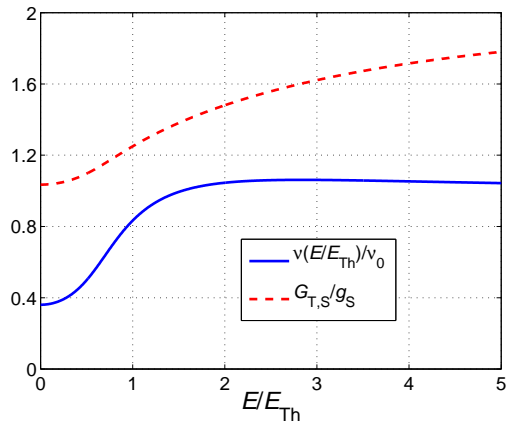}
  \caption{(Color online) Spectral properties of the cavity when the
    contact to the superconducting reservoir is metallic with
    $T_n^{(\text{S})}=1$ for all propagating channels and zero
    otherwise, and the contacts to the normal reservoirs are of tunnel
    type with $g_1/g_\text{S}=0.1$ and $g_2/g_\text{S}=0.3$. Solid
    line (blue) shows normalized density of states in the cavity,
    $\nu(E)/\nu_0$, and dashed line (red) shows
    $G_\text{T,S}/g_\text{S}$ which is the parameter that determines
    the Andreev reflection rate of quasiparticles in the cavity.}
  \label{fig:spectral-Sballistic}
\end{figure}

Systems where two normal metal terminals are connected to a cavity, in
which the cavity may itself be part of a larger superconducting
structure, can be studied in our model. The contact to the
superconductor could in this case be e.g. diffusive or metallic. In
this section, we assume tunnel contacts to the two normal metals, and
metallic contact to the superconductor. The metallic contact is
described by the transmission probabilities $T_n^{(\text{S})}=1$ for
all propagating channels and zero otherwise. We choose parameters
$g_1/g_\text{S}=0.1$ and $g_2/g_\text{S}=0.3$. The Thouless energy is
$E_\text{Th}=\hbar g_\text{S}/(2e^2\nu_0 V_\text{c})$. In Fig.
\ref{fig:spectral-Sballistic} we show the spectral properties of the
cavity. Below the Thouless energy, the quasiparticle density of states
is suppressed. At low energy, $G_\text{T,S}$ approaches the value in
the normal state. This is similar to the reentrant behavior in
diffusive systems.\cite{Stoof:prb96,chien:15356} With increasing
energy, the rate of Andreev reflection at the superconducting terminal
increases. The value of $G_\text{T,S}/g_\text{S}$ reaches its maximum
value 2 for $E/E_\text{Th}\approx 20$. This result agrees with the
Blonder-Tinkham-Klapwijk formula for high transmission probabilities,
valid when proximity effect is negligible.\cite{Blonder:prb82}

\begin{figure}[htbp]
  \includegraphics[scale=0.5]{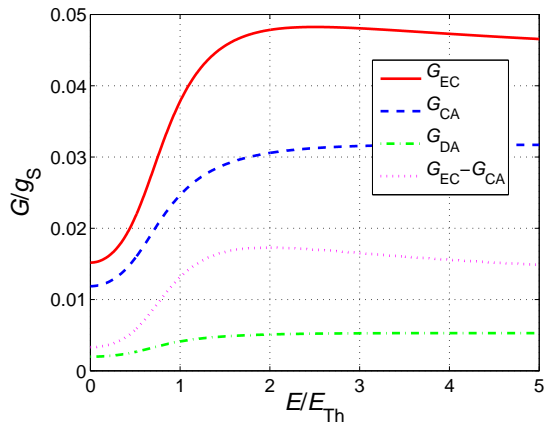}
  \caption{(Color online) Conductances for the various transport
    processes when the contact to the superconducting reservoir is
    metallic with $T_n^{(\text{S})}=1$ for all propagating channels
    and zero otherwise, and the contacts to the normal
    reservoirs are of tunnel type with $g_1/g_\text{S}=0.1$ and
    $g_2/g_\text{S}=0.3$. Solid line (red) $G_\text{EC}/g_\text{S}$,
    dashed line (blue) $G_\text{CA}/g_\text{S}$, dash-dot line (green)
    $G_\text{DA}/g_\text{S}$, and dotted line (purple)
    $(G_\text{EC}-G_\text{CA})/g_\text{S}$.}
  \label{fig:conductances-Sballistic}
\end{figure}

The conductances for this system are shown in Fig.
\ref{fig:conductances-Sballistic}. All conductances are suppressed
below the Thouless energy by the low density of states. Above the
Thouless energy, the increasing Andreev reflection rate leads to a
suppression of $G_\text{EC}$ and an enhancement of $G_\text{CA}$. The
overall behavior of the nonlocal conductances is determined by the
interplay of the dependence on the density of states and the Andreev
reflection rate. The two conductances associated with Andreev
reflection, $G_\text{CA}$ and $G_\text{DA}$, reach their maximum at
approximately $E/E_\text{Th}=4$ and then decrease slowly, reaching
their final value at $E/E_\text{Th}\approx 20$.

\subsection{\label{sec:Nballistic}Metallic contacts to N$_1$ and N$_2$,\\
  tunnel contact to S}
\begin{figure}[htbp]
  \includegraphics[scale=0.5]{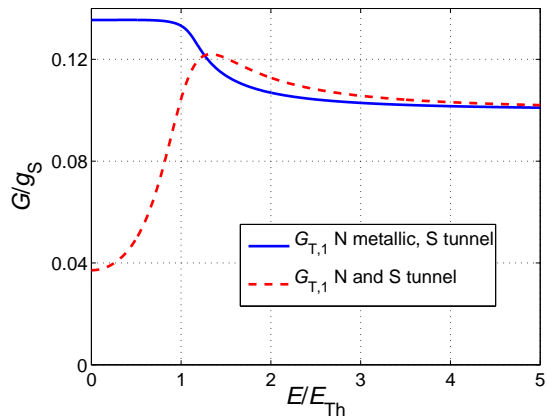}
  \caption{(Color online) Energy-dependence of conductance
    $G_\text{T,1}$ for the present case of metallic contacts to normal
    reservoirs and tunnel contact to the superconducting reservoir in
    solid line (blue), and for the case with only tunnel barriers
    studied in Sec.  \ref{sec:tunnel} in dashed line (red). In both
    cases we have put the parameters $g_1/g_\text{S}=0.1$, and
    $g_2/g_\text{S}=0.3$}
  \label{fig:compare}
\end{figure}

Let us now consider a system where the normal terminals N$_1$ and
N$_2$ are connected to the cavity through metallic contacts described
by $T_n^{(i)}=1$ for all propagating channels and zero otherwise. The
superconducting terminal is connected by a tunnel barrier of
conductance $g_\text{S}$, and the Thouless energy is
$E_\text{Th}=\hbar g_\text{S}/(2e^2\nu_0 V_\text{c})$. The spectral
properties of this system are very similar to the tunneling case in
Section \ref{sec:tunnel}, and we do not show them here. However, the
conductances between the normal terminals and the cavity are
qualitatively different in the present case. In Fig.
\ref{fig:compare} we show $G_\text{T,1}$ for both the tunneling case
in Section \ref{sec:tunnel}, and where the normal terminals are
connected by metallic contacts. At energies below the Thouless energy,
$G_\text{T,1}$ is in the tunneling case qualitatively similar to the
density of states. However, with metallic contacts to the normal
reservoirs, $G_{\text{T},1}$ is large at zero energy and decreases as
the energy increases beyond $E_\text{Th}$. At high energy,
$G_\text{T,1}$ is equal for the two cases and corresponds to the
result in the normal state since $g_1/g_\text{S}$ is the same for the
two curves. $G_\text{T,1}$ for a metallic contact is qualitatively
similar to the conductance of a metallic normal metal- superconductor
interface, except that in this case the ``superconductor'' is the
cavity which is under the influence of the proximity effect from S and
the relevant energy scale is $E_\text{Th}$ instead of $\Delta$.

\begin{figure}[htbp]
  \includegraphics[scale=0.5]{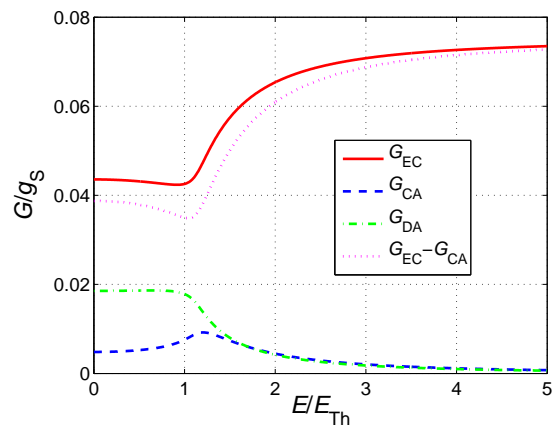}
  \caption{(Color online) Conductances for the various transport
    processes when the contacts to the normal reservoirs are of
    metallic type with $T_n^{(i)}=1$ where $i=1,2$ for all propagating
    channels and zero otherwise, and the contact to the
    superconducting reservoir is of tunnel type with conductance
    $g_\text{S}$. In this plot $g_1/g_\text{S}=0.1$ and
    $g_2/g_\text{S}=0.3$. Solid line (red) $G_\text{EC}/g_\text{S}$,
    dashed line (blue) $G_\text{CA}/g_\text{S}$, dash-dot line (green)
    $G_\text{DA}/g_\text{S}$, and dotted line (purple)
    $(G_\text{EC}-G_\text{CA})/g_\text{S}$.}
  \label{fig:conductances-Nballistic}
\end{figure}

The conductances for the present system are shown in Fig.
\ref{fig:conductances-Nballistic}.  A new feature here is a small dip
in $G_\text{EC}$ at $E=E_\text{Th}$ before the rapid increase above
$E_\text{Th}$. $G_\text{EC}$ is proportional to $G_\text{T,1}$ and
inversely proportional to $G_\text{T,S}$. The dip in $G_\text{EC}$ can
therefore be understood by the decreasing charge current conductance
($G_\text{T,1}$) and the peak in $G_\text{T,S}$ around
$E=E_\text{Th}$. At higher energy $G_\text{EC}$ increases as
$G_\text{T,S}$ vanishes. The dip in $G_\text{EC}$ leads to a larger
dip in $G_\text{EC}-G_\text{CA}$ since $G_\text{CA}$ increases with
increasing energy at $E<E_\text{Th}$. In the tunneling case of Section
\ref{sec:tunnel}, the increase of the Andreev conductance is
compensated by an increasing charge current conductance in
$G_\text{EC}-G_\text{CA}$, and there is no dip in the nonlocal
differential conductance. At high energy, $G_\text{CA}$ and
$G_\text{DA}$ vanish since the Andreev reflection rate vanishes.

\section{\label{sec:conclusion}Conclusion}                
In conclusion, we have studied nonlocal transport in a three terminal
device with two normal metal terminals and one superconducting
terminal. To this end we have applied the circuit theory of mesoscopic
transport. The connectors between the circuit elements are represented
by general expressions, relevant for a wide range of contacts.
Dephasing is taken into account, and is governed by the inverse dwell
time. This gives rise to an effective Thouless energy. For this model
we have calculated the conductance associated with crossed Andreev
reflection and electron cotunneling, as well as direct Andreev
reflection. We find that for any contacts, electron cotunneling is the
dominant nonlocal transport process. Results for various combinations
of contacts of tunnel and metallic type are shown, and this
demonstrates the strong dependence on the nature of the contacts. The
conductance of nonlocal transport has a strong dependence on the
Thouless energy.  This is because at these energies, dephasing leads
to a loss of induced superconducting correlations. Several of the
characteristics of our model have been observed in experiments,
although we do not make complete qualitative contact to experimental
data since electron cotunneling always dominates crossed Andreev
reflection. We suggest determining the conductance due to these
processes independently by carrying out additional energy transport
measurements.

\begin{acknowledgments}
  This work was supported in part by The Research Council of Norway
  through Grants No. 167498/V30, 162742/V00, 1534581/432, 1585181/143,
  1585471/431, RTN Spintronics, the Swiss NSF, the NCCR Nanoscience,
  the Deutsche Forschungsgemeinschaft through SFB 513, the
  Landesstiftung Baden-W\"{u}rttemberg through the Research Network
  "Functional Nanostructures", the National Science Foundation under
  Grant No.  PHY99-07949, and EU via project NMP2-CT-2003-505587
  'SFINx'.
\end{acknowledgments}

\begin{appendix}
  \section{\label{app:notation}Notation}
  Matrices in Nambu and Keldysh matrix space are denoted by a ``hat''
  ($\hat{M}$) and a ``bar'' ($\bar{M}$) respectively. The symbol used
  for a unit matrix is $\unitmatrix$, and the Pauli matrices are
  denoted $\hat{\sigma}_n$ and $\bar{\tau}_n$ in Nambu and Keldysh
  space where $n=1,2,3$. Compositions of matrices in Nambu and Keldysh
  space are formed by a direct product to make up 4$\times$4 matrices
  in Nambu-Keldysh matrix space, so that e.g.
  \begin{equation}
    \label{eq:2}
    \hat{\sigma}_3\bar{\tau}_1=\begin{pmatrix}0&0&1&0\\0&0&0&-1\\1&0&0&0\\0&-1&0&0\end{pmatrix}.
  \end{equation}
\end{appendix}

\bibliography{/home/gudrun/janpette/artikkel/fs}

\end{document}